\documentclass[12pt]{article}
\date{}

\title{Time and the laws of Nature}

\author{Alexey Mazur\thanks{E-mail: mazur@sol.ru}}

\begin{document}

\maketitle

\abstract{What is time? Why does it "flow" and why are we sure that it flows
from past towards future? Why is there such a gigantic distinction
between the Past of our world, which we believe to be fixed, and
the Future, which we consider undetermined? And this is in spite
of the fact that almost all physical laws are time-symmetric! Why
does previously undetermined "Future" acquire its steadiness as it
passes the moment called "the Present"? Perhaps, these questions
are partially answered in the considerations below.}

\vspace{1.5cm}
Let us retract from our particular Universe and think about what
kind of laws can in principle govern the development of universes.
As a universe, we will understand any world, even an imaginary
one, that can have different states and has laws of transition
from one state to another as "time" passes by. Generally speaking,
such laws can be probabilistic, i.e. they set the probabilities of
transition from one particular state to another.

Let us group these laws in categories on the basis of their
respect to time.

The first, and the most widespread in our world, category is the
category of bilaterally deterministic laws. "Bilaterally" means
that these laws determine unambiguously what was the precursor and
what will be the successor to a given state. (For the sake of
simplicity, we will assume time discrete.)

The second possible type of law embraces bilaterally
undeterministic (probabilistic) laws. In other words, starting
from a given present state, we can arrive, with a certain
probability, to a number of different states. The present state,
in its turn, can be the result of evolution of several possible
states in past, again with certain probabilities. (The case of
complete chaos falls precisely into this category.)

Laws of the third type, which we intend to discuss in detail, are
unilaterally deterministic laws. This means that such a law will
uniquely prescribe what a given state will evolve into. However,
the present state itself can originate from different states in
past.

At the first glance, one could think of laws of the fourth
category, i.e. laws that fix the ancestor to a given state but is
probabilistic in respect to its successors. However, since the
direction of time axis is subjective, this category is identical
to the third one, which will be discussed later in more detail.\\

{\em An example of the universe evolving in compliance with the
third type of laws is given by the famous Conway's Game of Life.
Let me briefly describe it. A 2D square grid is populated by a
colony of tips, each tip living in one cell. If a tip has two or
three neighbors, it survives and proceeds to the next generation.
Tips with zero or one neighbor die of "loneliness", while those
with more than three neighbors die of "overcrowding". New tips are
born in cells contiguous to exactly three living tips. The laws of
evolution in this world are more than simple, with every next
generation being unambiguously determined by the previous one; the
development of colonies can turn out very interesting.

However, it is plain to see that it is far from an easy task to
restore the previous state of a given one. Moreover, there can be
a multitude of them. Now think of a researcher who watches how
these states turn into one another but he watches this backward in
time. He is trying to determine the laws this evolution follows
but at best he can deduce only probabilistic distributions to
obtain this or that state from a given one.

What does it look similar to?}\\

Now I proceed to my principal statement: {\bf \large the laws of
our world are of unilaterally deterministic nature, if we choose
"from future to past" as the correct direction of the time axis.}

Apparently, the combination of unilaterally and bilaterally
deterministic laws results in the unilaterally deterministic
world, therefore the majority of the known laws does not determine
the general nature of our universe. Moreover, the majority of laws
are known to be CP-even and thus can be classified as laws of the
first or second category. (As a matter of fact, I am deeply
convinced that there are no laws of the second kind in our
universe. Otherwise, if there were several alternatives to past,
how could we remember the only one of them, known as "the Past"?)
The only exclusion I am aware of (I apologize for my possible
ignorance) is the weak interaction. Evidently, it is the weak
interaction that possesses the unilaterally deterministic nature
and, in absence of the competitors, determines the unilateral
nature of our universe. This explains the enormous difference
between the past and the future, which is clearly apparent to any
of us. We do remember our past, because it is uniquely determined
by the present. We know that millions of years ago the Earth was
inhabited by dinosaurs but no one has even a slightest idea who
will occupy the Earth million years from now. We can "remember"
and accumulate information because the past is strictly
determined. Almost entire (see below more on "almost") information
about the past is contained in the present. But not the
information on the future. If the laws of nature were bilaterally
deterministic, there would be no chance of any slightest
development (in our understanding of this word). Time would be
simply another dimension of the space, though quite a peculiar
one.\\

Now - several speculative remarks on why our Universe is what it
is and what conclusions can be drawn from the above discussion.

1. Of course, one might seem quite naive to tackle this "why"
question, but nonetheless I suggest doing this on the basis of
assumption that the world is designed in such a way that it
ensures the existence of OBSERVERS. In other worlds, even if they
exist, there is nobody to ask this question. So, starting from
universal CPT conservation, one can conclude that in the world
built of antimatter time will flow in the opposite direction. I.e.
in this world the future is fixed, while the past can have
alternatives. So, the well known problem "how to determine by
means of radiowave communication whether you are talking to a
creature built of matter or antimatter" no longer arises. Indeed,
such a contact is just impossible, since the same signal will be
considered emitted by both parties (by an antenna - at the one
end, by synchronously distinct atoms towards the antenna - at the
other). The information exchange is simply impossible. But
nevermind, there seem to be no problems with antimatter creatures
in our world by the virtue of almost absolute absence of the
antimatter itself. But since the antimatter has "time-opposite"
nature, it introduces indeterminacy into our past. Obviously, it
is matter-to-antimatter relation that dictates the relation
between indeterminacy in past and future. These two forms of
disproportion - between matter and antimatter content of our
universe, and between determinacy in past and future - are not
only connected, but also necessitate one another.

2. One should undertake a search for some relatively simple laws
that the weak interaction follows in the "backward time".

3. If there exist only finite number of states of universe, there
must also exist special states, called "Gardens of Eden", that
have (in the backward time) successors but do not have any
precursor (if some states have more than one precursors, then
there are no precursors left for some of the other). For the usual
direction of the time axis it means that there is NO SINGLE state
that our present state can evolve into. That is, "Garden of Eden"
for us is the apocalypse in its direct and literal meaning. In the
Game of Life such "Gardens of Eden" have been found. This problem
is however removed in the case of infinite number of states. (An
example: a universe, whose states are real numbers from 0 to 1.
The evolution law: take the infinite decimal fraction that
represents this number and erase the first decimal digit. The
resulting number represents the next state of the universe. It is
clear that every state of this universe has ten precursors.)

4. The model of unilaterally deterministic universe removes the
contradiction between "God does not play dice" and the freedom of
will, obviously given to a human being. If one wishes, one can
think that God has created a finite (finite for us) state of the
universe and the laws of universe evolution in the "backward
time". And now He is surprised to find people who think that
everything is evolving in the opposite time.

5. The choice of time axis direction is subjective. \\ If we
introduce another axis, arrange all states of the universe along
it, and undertake an unbiased analysis, we will find the
following. At each moment of time $t$ creatures dwelling in the
Universe are convinced that the moment of time $t-1$ was in past.
What is the ground for this belief? Only the fact that creatures
possess significantly richer information about this moment than
about moment $t+1$. And at the moment $t+1$ they "remember" that
"some time ago", at moment $t$, they have just learn about this
moment $t$. This tempts them to think that time is directed from
past to future. But I think that as far as the physical laws are
concerned, the correct choice of time axis direction should be
exactly opposite.
\end{document}